\documentclass{JINST}

\title{GARLIC: GAmma Reconstruction at a LInear Collider experiment}

\author{Daniel Jeans\thanks{Corresponding author.}, 
  Jean-Claude Brient, Marcel Reinhard\\
 Laboratoire Leprince-Ringuet - \'Ecole polytechnique, CNRS/IN2P3 \\ 
  Palaiseau, France \\ E-mail: \email{daniel.jeans@llr.in2p3.fr}
}

\abstract{The precise measurement of hadronic jet energy is crucial to maximise the physics
reach of a future Linear Collider. An important ingredient required to achieve this
is the efficient identification of photons within hadronic showers.
On configuration of the ILD detector concept employs a highly granular silicon-tungsten 
sampling calorimeter to 
identify and measure photons, and the GARLIC algorithm described in this paper has been 
developed to identify photons in such a calorimeter.
We describe the algorithm and characterise its performance using events fully simulated
in a model of the ILD detector.}

\begin{document}


\section{Introduction}
\label{intro}

A linear $e^+e^-$ collider (LC) with a center of mass energy at the TeV scale is foreseen for the 
study of high energy particle physics in parallel with and beyond the LHC era~\cite{ILCRDR}.
The identification of multi-boson events will be of major importance for the physics output of this machine, and 
consequently the ability to distinguish $Z$, $W^{\pm}$ and Higgs bosons will be crucial. 
Since the hadronic branching fractions of these bosons is large, 
a maximal use of the luminosity must include the use of bosons decaying to jets. A key test
of an experiment's performance is therefore the resolution with which it can measure the 
invariant mass of multi-jet systems.

The identification and measurement of fully hadronic $W^+W^-$ and $ZZ$ decays has already been performed at
the LEP experiments \cite{LEP2_ewcomb}.
However these analyses required the use of kinematical fitting to improve the di-jet mass resolution. 
This approach is not always possible for physics analyses at a LC, in particular events with more than a single
neutrino in the final state, for example the important processes 
$e^+ e^- \to W^+ W^- \nu \overline{\nu}$, $ZZ\nu \overline{\nu}$, and $H\nu \overline{\nu}$. 
In these cases the precision of the jet energy measurement is the crucial parameter for detector optimization.

The jet energy resolution required to achieve sufficient boson separation has been extensively studied 
(e.g. \cite{tesla_physics}).
It is now generally agreed that the LC physics program can be realised with a resolution on jet energy
of ${\rm \Delta E / E = 3\sim4\%}$~\cite{ILDloi}\cite{SiDloi}, 
significantly better than the resolutions achieved by the LEP experiments $\sigma(E) \sim (60-90\%) \sqrt{E}$
\cite{ALEPHperformance}\cite{DELPHIperformance}\cite{L3performance}\cite{OPALperformance}.
Several approaches to attaining this energy resolution are currently under study. 
The most promising is an event reconstruction technique known as Particle Flow (PFA) \cite{pfa_jcb_hv}.
This technique is based on the individual reconstruction of all final state particles.  
Considering the average energy content of a jet with 65\% carried by charged particles, 26\% by photons and 
about 9\% by neutral hadrons \cite{jetcompos}, it is natural to use a magnetic spectrometer to estimate the 
energy of charged particles since these detectors typically have a momentum resolution much better than
the energy resolution of a calorimeter. 
A calorimeter designed for PFA is therefore dedicated to measuring the energy of only neutral particles, and it must
be able to disentangle the energy deposits of neutral and charged particles. 
The minimisation of the ``confusion'', 
the incorrect assignment of charged energy deposits to neutral and vice versa, is the central
challenge to the particle flow approach. 
A detector with highly segmented calorimeter readout is, a priori
the most suitable for the application of PFA.

In this paper, we first briefly describe the International Large Detector detector concept, emphasising its
highly-granular electromagnetic calorimeter, and then discuss the role of photon identification within the 
particle flow method. In section \ref{algo} we describe the GARLIC algorithm, and then measure its
performance in fully simulated hadronic jet events.

\section{The International Large Detector}

The International Large Detector (ILD) is designed to reconstruct events produced at a LC 
using the particle flow technique.
There are several technological options for some of ILD's sub-detectors. The options simulated for
the analyses reported in this paper are those of the ILD\_00 model.
This section gives a brief overview of this model, full details of which can be found in \cite{ILDloi}.

ILD is a general purpose particle physics detector providing close to $4\pi$ solid angle coverage.
It is conceived as an approximately cylindrical central barrel section closed by two endcaps.
Starting from the interaction point, which lies within a beryllium beam pipe of radius 15 mm, the detector consists
of a five-layer silicon vertex detector complemented by silicon tracking disks in the
forward region, followed by a large time projection chamber (TPC). Silicon strip detectors
are placed immediately before and after the TPC. This gives a very precise tracking system
with excellent pattern recognition capabilities.
The tracking detectors are followed by highly granular sampling calorimeters. 
First is a thirty-layer silicon-tungsten electromagnetic calorimeter (ECAL), with
a transverse segmentation of $5 \times 5 {\rm mm^2}$, followed by
a scintillator-steel hadronic calorimeter with 48 layers and a 
transverse segmentation of $3 \times 3 {\rm cm^2}$.
The calorimeters are followed by a superconducting solenoid which provides a highly uniform field
of 3.5 T in the central part of the detector. 
Endcaps consisting of similar calorimeter systems close the central barrel structure.
The steel return yoke is instrumented with muon chambers.

The ILD\_00 detector model has been simulated in the Geant4-based MOKKA simulation framework~\cite{Geant4}~\cite{MOKKA}.

\section{Electromagnetic calorimeter}

The identification of photons uses mostly information from the electromagnetic calorimeter.
The barrel section of the ECAL is constructed of eight ``staves'' forming an octagonal tube,
with each stave made up of five identical ``modules'' along the beam axis.
Each module consists of a carbon-fibre/tungsten support structure which supports the thirty
silicon detection layers. The total thickness of the ECAL is around 20 cm, corresponding to $24 {\rm X_0}$ and $1 \lambda_I$,
and has an effective Moli\`ere radius of around 19~mm.
Each module consists of two ``stacks'' with different
thicknesses of tungsten layers: the first twenty absorber plates have a thickness of 2.1~mm ($\rm 0.6~X_0$), the remaining 
nine plates are 4.2~mm thick.
The silicon sensors have a size of $9\times9 {\rm cm^2}$, and are segmented into
$5 \times 5 {\rm mm^2}$ detection cells in all detector layers. 
Non-instrumented regions are simulated,
with a 1~mm region between the active areas of adjacent sensors, 
an additional 1~mm gap corresponding to the supporting walls within a module, 
and an additional 2~mm gap between adjacent modules.
The two endcap ECALs are split into quarters, each of which is composed of three
modules with a similar structure (but different overall shape) to the barrel modules.

Each detector cell is read out using electronics with a high dynamic range, 
allowing the measurement of signals ranging from 0.5 MIP to 2500 MIPs
(where a MIP is the most probable signal produced by the passage of a 
Minimum Ionising Particle).
Typical noise in the electronics has a width of around 10\% of a MIP.
Hits with an energy of less that 0.5 MIP are suppressed.

\section{Particle flow and photon identification}

Particle flow relies on the ability to distinguish the deposits of 
the charged and neutral particles in a hadronic jet by topological means. 
The momentum of charged particles (measured in the tracking system) is used to estimate
the charged jet energy, while the neutral energy is estimated by considering the 
energy deposited in the calorimeters.
The significant energy carried by the photonic jet component (predominantly produced in neutral pion decay) 
must be efficiently and cleanly identified in order to prevent over- or under-counting of energy, 
which would lead to an increase of the confusion term in the jet energy resolution.
Due to the compact and distinctive form of their energy deposit in the ECAL, the identification
of photons is typically the first step of particle flow reconstruction in the calorimeters.
Once photons are identified, the remainder of the event is analysed to reconstruct
the charged and neutral hadron components.

The single photon energy resolution of the ECAL must be sufficiently good that it does not
dominate the final jet energy resolution. This is typically satisfied with a moderate
energy resolution of $\delta E / E \sim (15\rightarrow20\%)/\sqrt{E [GeV]}$.

Of existing particle flow algorithms (e.g. \cite{Pandora}\cite{wolf_pfa}\cite{iowa_pfa}), 
PandoraPFA \cite{Pandora} has already shown that the aim of a 3\% jet energy resolution is realistic, 
and can be already achieved over a wide range of jet energies. 
These algorithms use generic clustering algorithms 
(typically based on nearest neighbours or directional cone clustering) to treat hits in 
the entire calorimeter system. 
Studies \cite{wilson_pi0} have shown that the reconstruction and constrained 
kinematic fitting of $\pi^0$ with a hadronic jet can give further improvements to jet energy resolution.

The GARLIC algorithm described in this paper is a clustering
algorithm dedicated to clusters produced by photons, using the known characteristics of
photon showers to inform the clustering process. The application of such a particle-specific 
approach has the potential to improve the performance of particle flow algorithms using 
generic clustering procedures.

\section{GARLIC algorithm}
\label{algo}

GARLIC is an algorithm designed to identify photons produced at the interaction point.
It makes use of the characteristic form of electromagnetic showers in the ECAL: the
narrow width of the shower (the Moli\`ere radius of this ECAL is around 2cm) and its
longitudinal profile, both of which are well measured in the highly granular
calorimeter proposed for ILD.

For use in a particle flow algorithm, photon finding must be efficient, and have as small
as possible contamination from charged particles. Such a contamination gives rise to the ``confusion''
term, typically the limiting factor to particle flow. Contamination from neutral hadrons in the
photon sample is relatively benign for the measurement of jet energy, since it does not contribute to the 
confusion term.

GARLIC is implemented as a Marlin c++ processor within the ilcsoft framework \cite{ilcsoft},
which is available for download via svn \cite{GARLICcode}.
It is designed to be executed as part of a complete reconstruction chain, 
using as input collections of reconstructed tracks 
(in this case produced by the FullLDCTracking Marlin processor \cite{LDCtracking}) 
and calibrated calorimeter hits (from the NewLDCCaloDigi processor of ilcsoft, with a 
hit energy threshold of 60\% of the most probable minimum ionising 
particle (MIP) energy deposition).

The algorithm considers only calorimeter hits at some distance to the extrapolations of reconstructed tracks. 
Using these hits, the algorithm identifies ``seeds'' using hits in the early layers of the calorimeter.
If these seeds pass some simple criteria, ``cores'' are built up, corresponding to the high energy
central region of EM showers. ``Clusters'' are then built up around these cores, attaching lower energy hits
in the halo of the EM shower.
Artificial neural networks are trained to decide if these clusters are photon-like, by considering 
several reconstructed cluster properties.

The following outlines the main steps of the GARLIC algorithm in more detail.

\subsection{Definitions}

\begin{itemize}
\item{Pseudo-layer}
The algorithm makes extensive use of the depth of calorimeter hits inside the calorimeter.
A hit's layer is a reasonable estimator of this depth within the ECAL
except in the transition regions between modules in $\phi$ and between the endcap and barrel. 
In these regions, 
the concept of a ``pseudo-layer'' is introduced (as in \cite{Pandora}), which
better approximates the true depth within the calorimeter in the transition regions.

\item{Seed}
The seed is an estimate of the photon impact point on the calorimeter, from which a 
core is built. A seed is a position in space with an associated direction.

\item{Core}
A core consists of the central high energy region of an EM shower.
A core is constructed from the hits within a cylinder of radius 1.5 times the cell size,
whose axis is defined by the position and direction of a seed. 
Hits in the first ECAL stack within this cylinder are added to the core until 
a gap of three successive empty layers are found. If no such gap is found, and at least
five first stack hits have been found in the first $10 {\rm X_0}$ of the ECAL, hits from the second stack are added
until two successive empty layers are found.

\item{Cluster}
A cluster is built around a core, in order to collect almost all hits produced by the photon. 
Hits within a certain distance of a core's hits are added to the cluster. 
This procedure is typically repeated a number of times, considering 
the distance to hits added to the cluster in the previous iteration.
\end{itemize}

\subsection{Electron veto}

A rather simple electron veto has been developed to identify hits deposited by electrons.
Charged tracks are extrapolated to the front face of the ECAL, where their position and
direction are used to define a seed. A core is built from this seed. All hits within 
40 mm of the core are considered for further cluster building, with five iterations
with a clustering distance 1.9 times the cell size.

If the resulting cluster has an energy consistent with the momentum of the track used to seed it,
it is flagged as an electron and its hits ignored for further photon cluster finding.

\subsection{Charged hit veto}

To remove ECAL hits produced by non-showering charged particles, reconstructed tracks 
are projected into the ECAL, and
ECAL hits closer than 10 mm to a projected track are removed and not further considered by the algorithm. 

\subsection{Pre-clustering}

Remaining hits are then split into pre-clusters using a simple nearest-neighbour clustering with
a clustering distance of $50 {\rm mm}$. Each pre-cluster 
containing at least 5 hits is independently considered by the GARLIC algorithm.

\subsection{Clustering}

\begin{enumerate}

 \item{Seed-finding.}
Each pre-cluster is analysed to identify cluster seeds. For this purpose, only ECAL hits
in the first twelve ECAL pseudo-layers (corresponding to ${\rm \sim7 X_0}$) with an energy of
at least 2.5 MIPs are considered. A line is defined between the pre-cluster's centre-of-gravity and 
the interaction point (IP), 
and a plane perpendicular to this line and passing through the line's intersection
with the ECAL front surface is considered. A two dimensional histogram 
(with a binning similar to the ECAL cell size) is filled with the energy-weighted projections
of precluster hits onto this plane. The highest energy bin of the histogram is identified,
and the average position of the 3x3 bins centred on it is calculated. Bins within $15 {\rm mm}$ (a little less than
the Moli\`ere radius)
of this position are grouped, and their average position used to define a cluster seed. This procedure 
provides a relatively stable position estimate for photon clusters, whether they hit the center or towards the 
edge of a detector cell.
The procedure is applied iteratively, starting with the highest energy unassociated bin.
The direction associated to these seeds is that of a line connecting it to the IP.
Seeds with less than 2 hits are rejected.

\item{Core building.}
These seeds are then used to build shower cores, as described above.
In the barrel-endcap overlap region, a wider cylinder is used in the endcap to account for the spreading of the 
electromagnetic shower in the barrel-endcap gap.

\item{Iterative clustering.}
A cluster is then built around this core, with five iterations using a clustering distance of 1.4 times the
cell size. In order to minimise the contribution from charged particles, if the centre-of-gravity of a cluster 
is very close to the extrapolation of a charged track (closer than 1.5 times the cell size) it is rejected. 
A cluster is also rejected if its energy is less than 150 MeV, since a clean
identification of photon clusters at such low energies seems to be extremely difficult 
with such an ECAL.


\item{Cluster merging}
A search is made for cases in which a single photon has been reconstructed as several clusters.
The number of pseudo-layers in which the smallest distance between the hits of two different clusters
is smaller than 1.5 times the cell distance is calculated. If this represents more than 75\% of the 
layers containing hits of both clusters, the two clusters are merged.

\subsection{Neural Network Verification}

Neural networks (NN) have been trained to distinguish GARLIC clusters originating from
photons and those produced by other particles.
The network was trained using the Multi-Layer Perceptron (MLP) method of the TMVA package~\cite{TMVA} using as 
training samples a mixture of two and four quark events at a centre of mass energy of 500 GeV, 
fully simulated in the ILD\_00 detector model. 

The GARLIC clusters identified in these events were split into six classes
according to their energy (with energy boundaries at 0.2, 0.5, 1, 3, and 10 GeV) and according
to the distance between the cluster projection and the nearest track extrapolation
onto the ECAL front face (closer and further than 100 mm). 
A NN was trained in each of these twelve cluster classes.
In the case of clusters close to a track, the network was trained to distinguish between 
clusters produced by generated photons which did not convert before reaching the ECAL, and
clusters produced by charged pions. For clusters far from a track, the networks were trained to
distinguish photon clusters from those not produced by photons or electrons and positrons.

Networks for clusters far from a track were trained with seven input observables:
\begin{itemize}
\item ``pointing angle'': angle between the cluster axis and the line between the interaction point (IP) and cluster centre of gravity (CoG). The vast majority of photons in jets originate at the IP.
\item ``mean depth'': energy-weighted mean depth of cluster hits. This observable is sensitive to the
typical longitudinal profile of EM showers.
\item ``energy fraction (5-10)'': fraction of cluster between $5$ and $10\ X_0$ after the cluster start. This is also
sensitive to the longitudinal shower profile.
\item ``mean hit energy'': mean of the cluster's hit energy distribution. EM showers also have a characteristic
distribution of hit energy.
\item ``RMS/mean hit energy'': ratio of the RMS of the hit energy distribution to its mean.
\item ``fractal dimension'' (FD) of the cluster: the log-ratio of the number of hits in the cluster (${\rm N_1}$) 
with the number of hits when cells are grouped into larger cells of 4x4 cells (${\rm N_4}$): 
${\rm FD = log_{10}(N_4/N_1)/log_{10}(4)}$. This measures how dense the shower is. EM showers are typically more
dense than hadronic showers.
\item ``minimum transverse RMS'': cluster hits are projected onto the front ECAL face along the direction between the cluster CoG and the IP. This 2-d projection is treated as an elliptical distribution, and the major and minor axes identified.
The 2-d projection is then further projected onto the minor axis, and the RMS of this projection is calculated.
This observable is useful because of the narrow shape of EM showers.
This particular construction has the advantage that it is quite insensitive to the case where 
two nearby photons are merged into a single cluster, which can distort the transverse shower shape.
\end{itemize}
In the case of clusters with a nearby track, two additional observables were used in addition to the above, 
by considering the track which intersects the ECAL front surface closest to the cluster. These help to distinguish
cases where a cluster is either independent of or created by the nearby track.
\begin{itemize}
\item ``track distance'': the distance between cluster projection onto the ECAL front face and the closest track extrapolation.
\item ``track angle'': the angle between the cluster direction and the direction of the nearest track at the ECAL front face.
\end{itemize}

The distribution of some of these observables are shown in figure \ref{fig:clusterParams}, and representative
output spectra from two neural network trainings are shown in figure \ref{fig:NNout}. Since the
NNs give continuous output distributions, the cut value used to decide if a cluster is photon-like
can be varied by the user to give a more efficient or more pure selection of photons. In the remainder of this
paper, we apply the cut at 0.5, which corresponds to the point at which a cluster has equal probability
to be a true photon or fake cluster in the event samples used to train the networks.

\begin{figure}
\center
\includegraphics[width=0.8\textwidth,keepaspectratio]{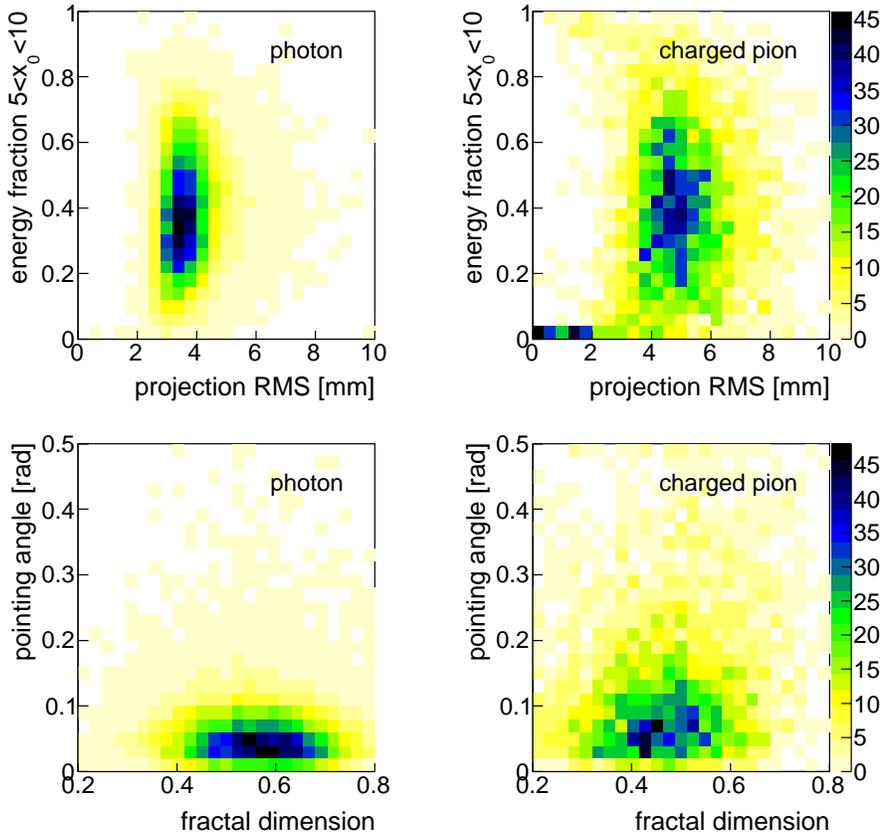}
\caption{Distribution of some observables used to train the Neural Networks.
``energy fraction (5-10)'' vs. ``minimum transverse RMS'' (upper plots) and
``pointing angle'' vs. ``fractal dimension'' (lower plots), 
for photon (left) and charged pion (right) clusters with an energy between 1 and 3~GeV found in 
4-quark events at 500 GeV. See text for the definition of these observables.}
\label{fig:clusterParams}
\end{figure}

\begin{figure}
\center
\includegraphics[width=0.48\textwidth,keepaspectratio]{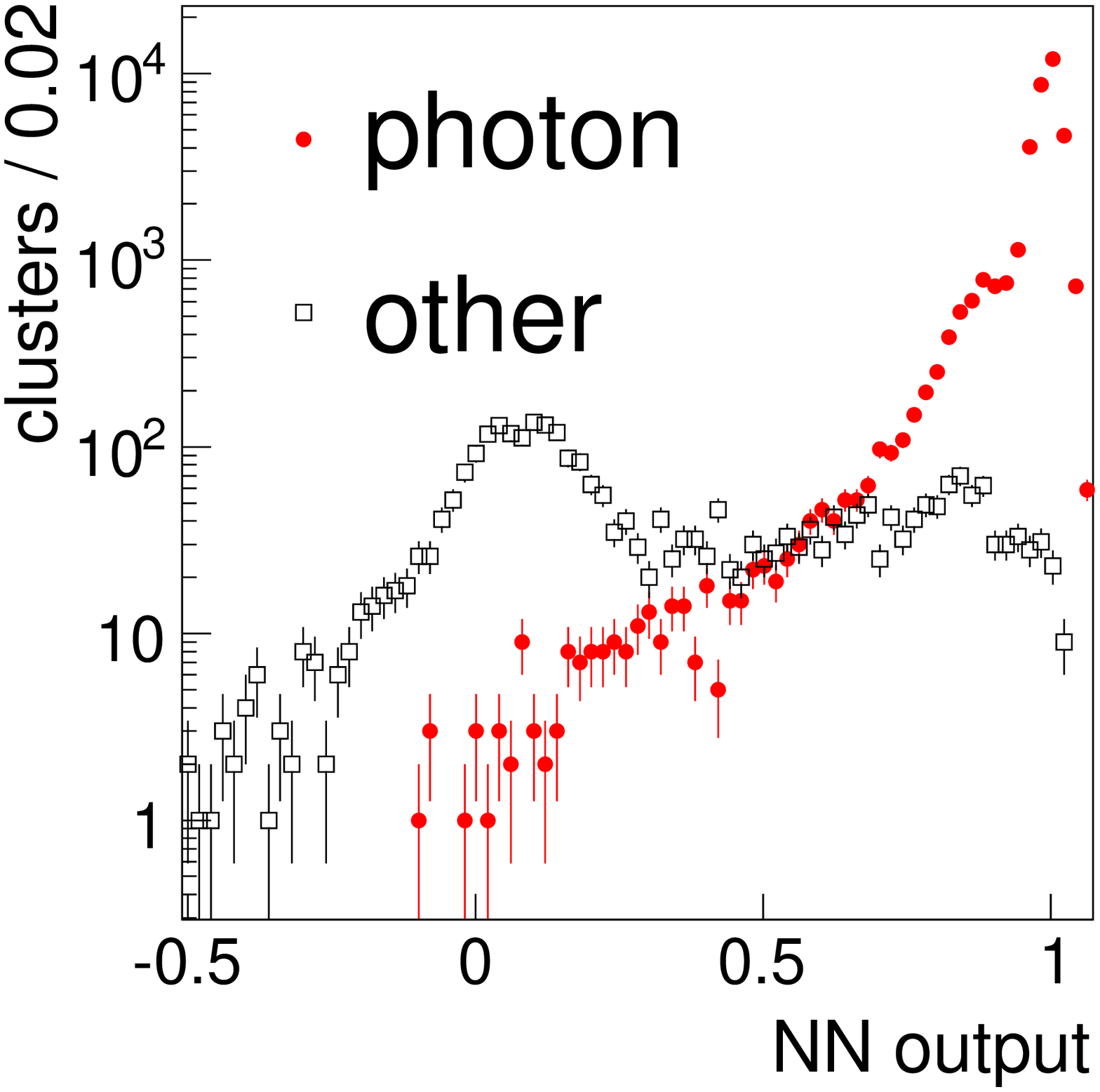}
\includegraphics[width=0.48\textwidth,keepaspectratio]{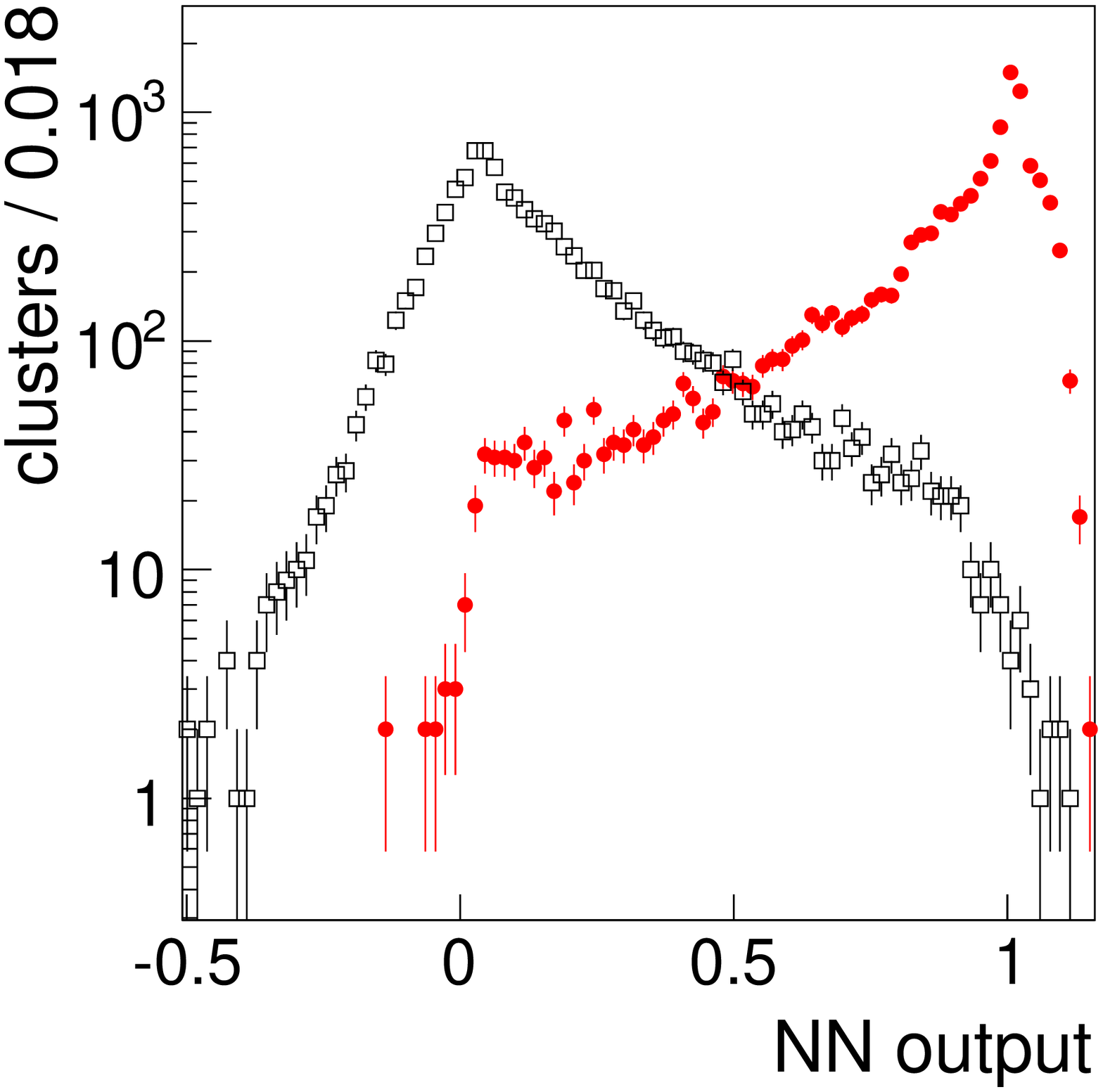}
\caption{Output of NN for clusters between 1 and 3 GeV, far (left) and near (right) from an extrapolated track.}
\label{fig:NNout}
\end{figure}

\end{enumerate}

%
%

\section{Performance}

To characterise the performance of the GARLIC algorithm, its performance 
was measured in samples of four quark events generated at a centre-of-mass energy of 500 GeV.
The samples used for performance measurement were independent of those used to train
the networks.

\begin{figure}
\includegraphics[width=0.45\textwidth,keepaspectratio]{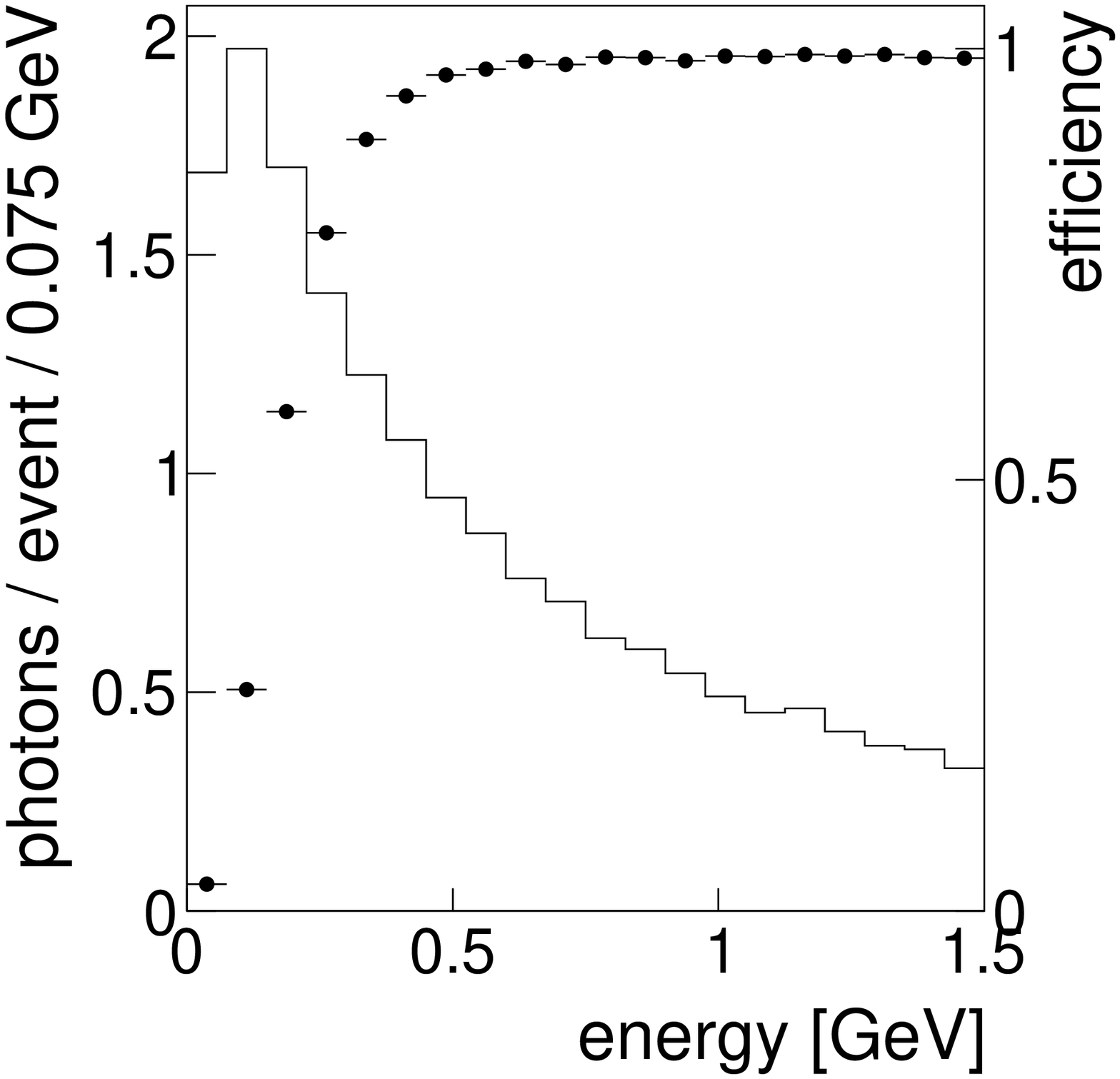}
\includegraphics[width=0.45\textwidth,keepaspectratio]{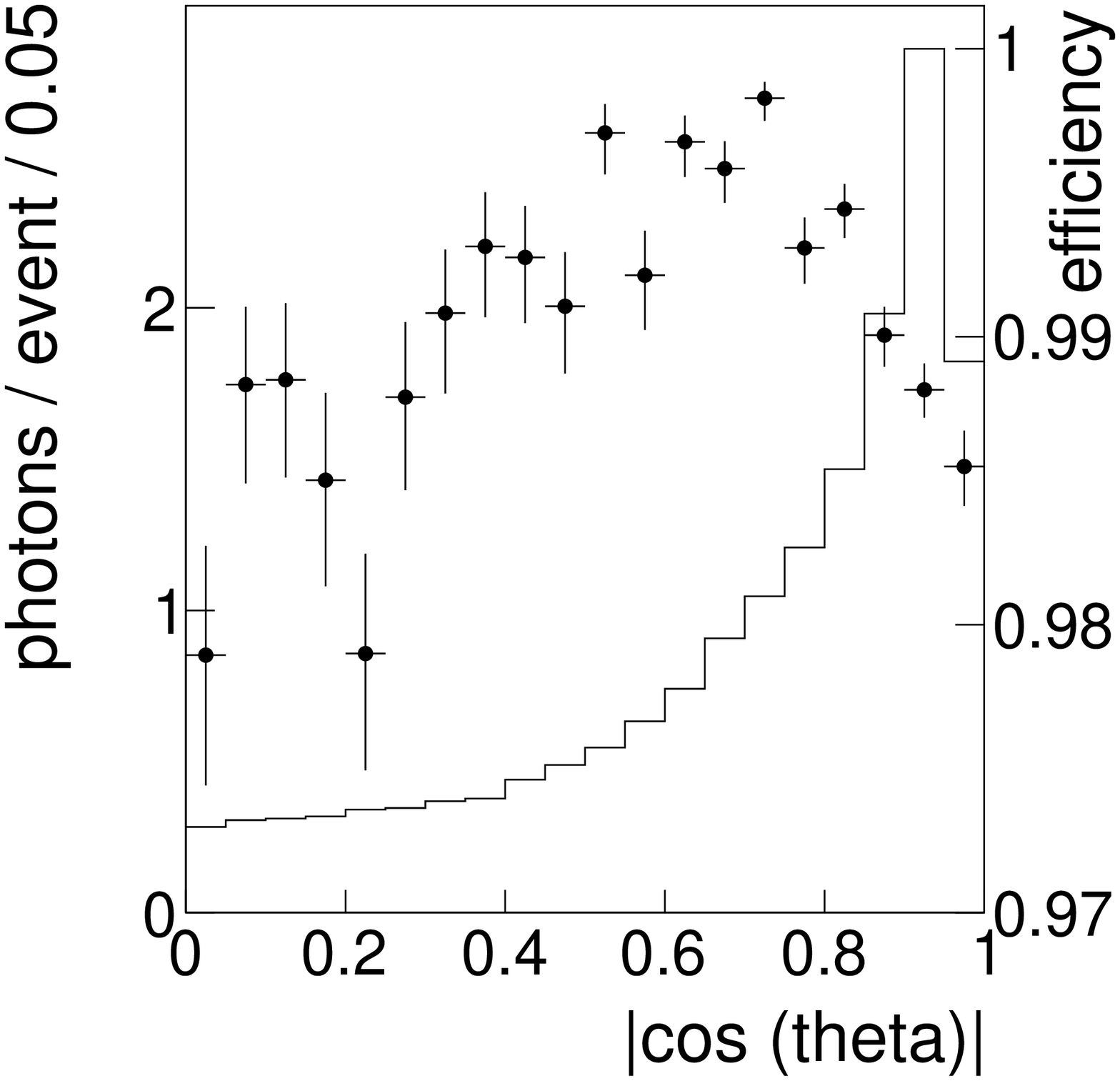} \\
\includegraphics[width=0.45\textwidth,keepaspectratio]{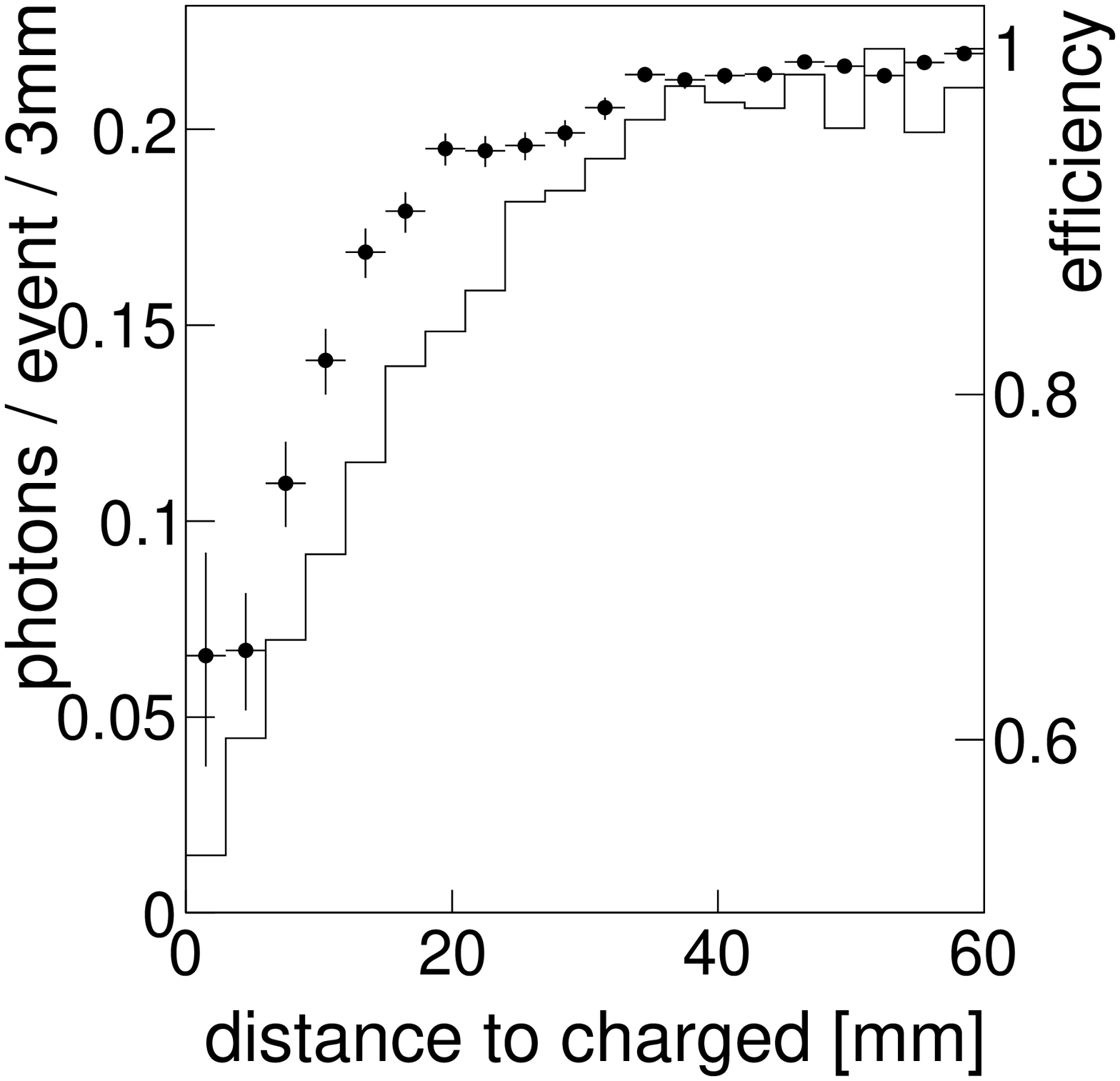}
\includegraphics[width=0.45\textwidth,keepaspectratio]{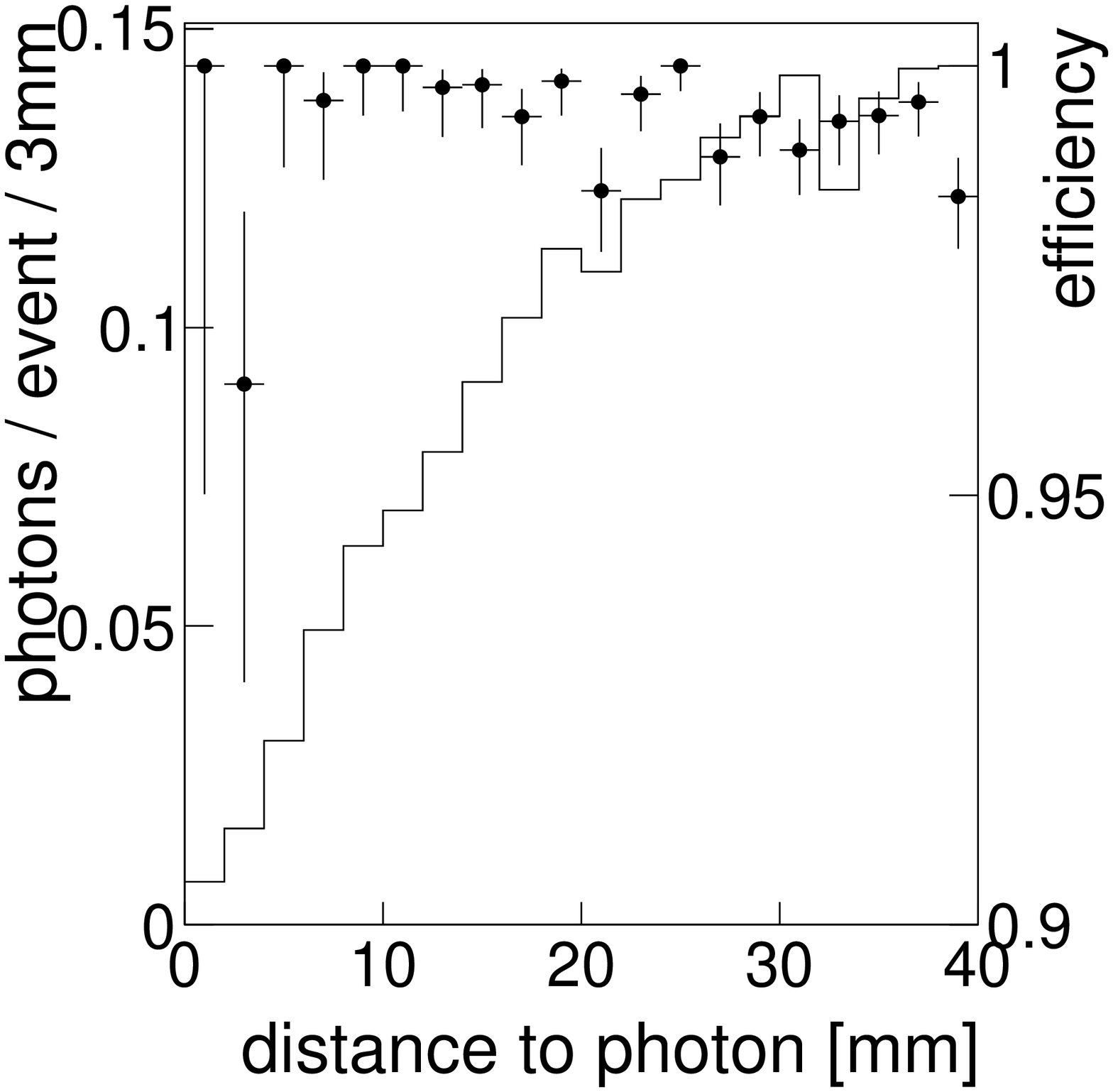}
\caption{Garlic efficiency for photons in four quark events generated at 500 GeV, 
as a function of various photon properties.
The histogram in each plot shows the distribution of photons in these variables, 
while the points with error bars show the estimated efficiency as a function of these variables.
In all plots other than those for the variables concerned,
the photon is required to have an energy of at least 500 MeV and be separated by at least 40 mm from the 
nearest charged particle extrapolation at the ECAL front face.
}
\label{fig:garlic_uddu_500}
\end{figure}

To measure the efficiency of the algorithm, generated photons which have not converted before reaching the ECAL 
are considered. The fraction of these photons which are within 0.01 radians (corresponding approximately to a 
Moli\`ere radius at the ECAL) 
of the closest reconstructed cluster CoG are considered as being successfully reconstructed. This efficiency
is shown in figure \ref{fig:garlic_uddu_500} as a function of several properties of the generated photon:
its energy, polar angle, and the distances to the nearest charged particle and nearest photon at the ECAL front face.
Superposed are the distribution of photons in the same four-jet events.
Except in the case of the energy graph, photons are required to
have an energy of at least 0.5 GeV. For all except the ``distance to charged particle'' plot, 
photons are required to be at least 40 mm from the nearest charged particle extrapolation at the front face of the ECAL.

The efficiency depends on the photon energy, with the efficiency dropping for photons below around 1 GeV. Above this energy,
the efficiency is around 99\%, it is around 80\% by 0.3 GeV and drops to 50\% by 0.2 GeV. 
This efficiency drop at low energy is largely due to the requirement that a cluster have an energy of at least 150 MeV.

When photons enter the ECAL close to a charged particle, the efficiency also drops for distances below around 40 mm. 
At 20 mm the efficiency is around 90\%, and drops further to around 60\% when the distance goes to zero. This drop
in efficiency is due largely to the NN selection, since when the cluster is in close
proximity to a track, the NN favours the hypothesis that it is due to a charged particle.

The dependence of the efficiency on other properties of the photon and its environment are much less strong. 
These is a modest dependence on polar angle, with variation between 98\% and around 99.6\%.
There is no discernible decrease in efficiency seen if the photon enters the ECAL close to a second photon.

The average efficiency for photons with an energy above 0.5 GeV and isolated by at least 40 mm from the nearest charged
particle is above 99\%.

To investigate the purity of selected GARLIC clusters, the particle which made the largest contribution to its energy was 
considered. If this particle was a photon either at the entry of the ECAL, or at the generator level 
(before any detector simulation) the cluster was considered correctly identified. 
This includes in the ``correctly identified'' sample cases where a photon has converted before reaching the ECAL 
(typically in the field cage or endplate of the TPC or the silicon SET/ETD trackers \cite{ILDloi} placed immediately before the ECAL), or
where the cluster is due to a brehmsstrahlung photon produced by a charged particle during its passage through 
the detector material in front of the ECAL. Clusters not falling into this category were classed according to the 
nature of the particle at the entry of the ECAL which gave the largest contribution to the cluster energy.

\begin{figure}
\center
\includegraphics[width=0.48\textwidth,keepaspectratio]{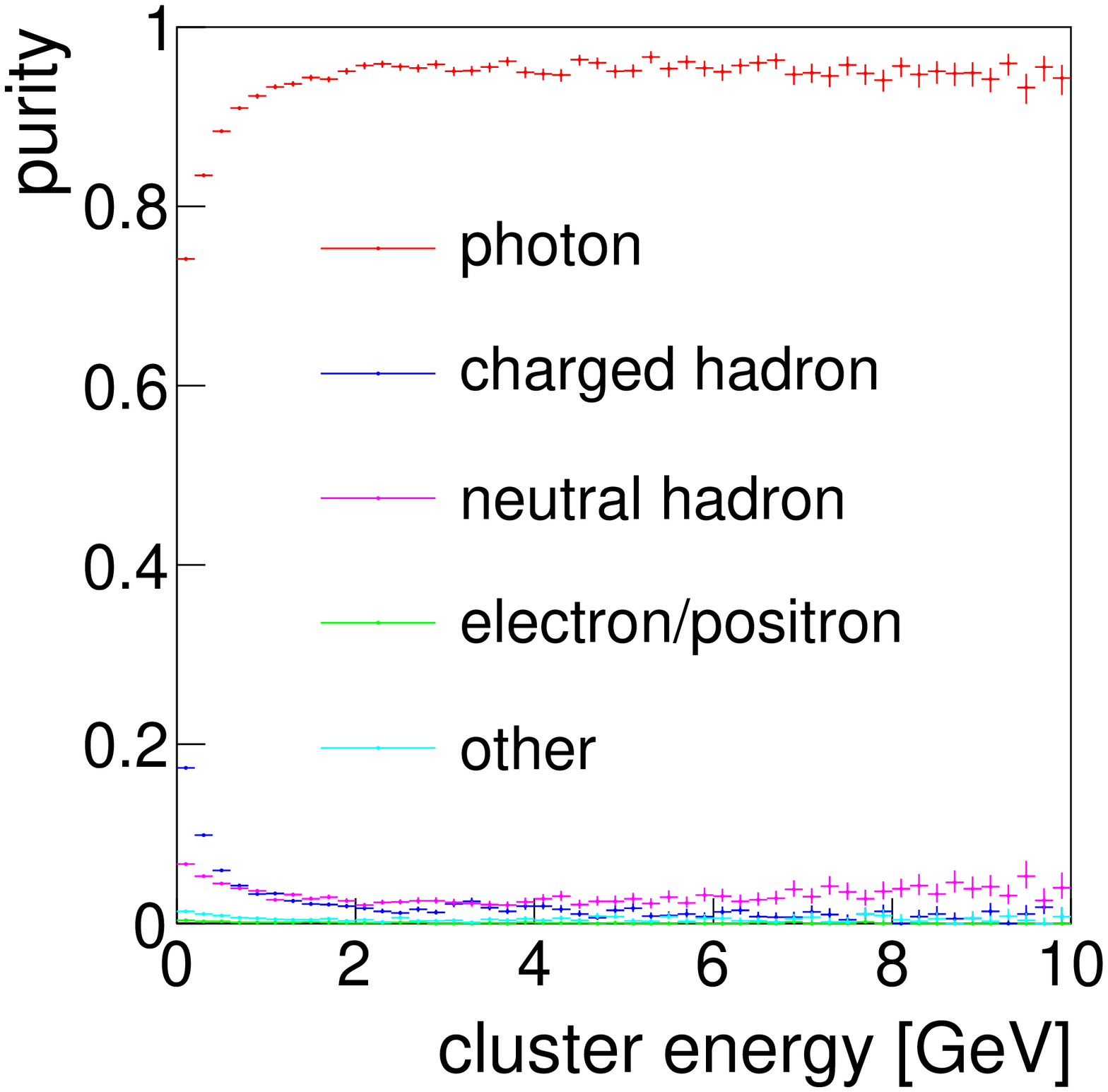}
\includegraphics[width=0.48\textwidth,keepaspectratio]{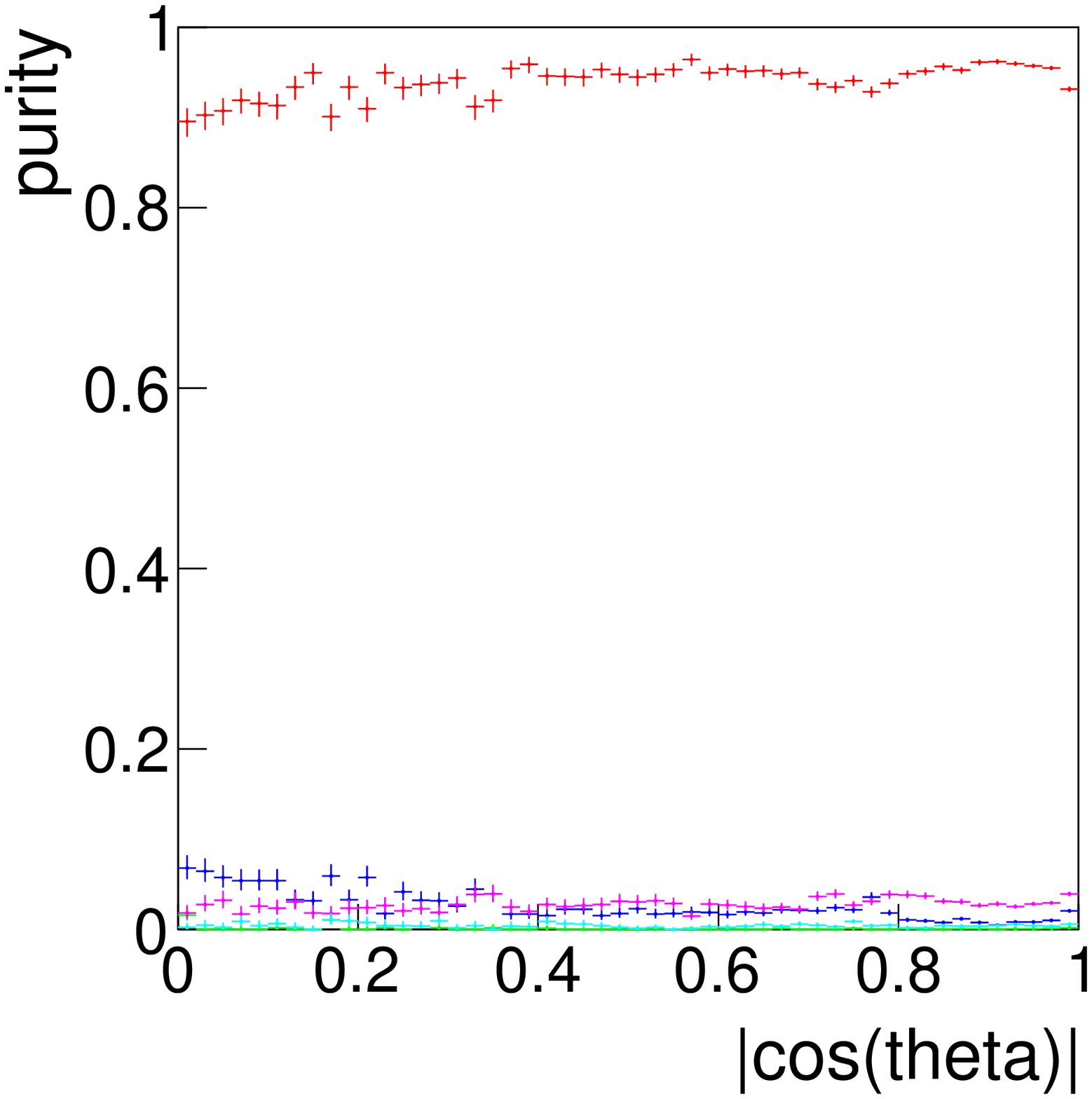}
\caption{Leading contributing particle of selected GARLIC clusters, as a function of energy and polar angle.
In the case of the polar angle plot, clusters were required to have an energy of at least 1 GeV.}
\label{fig:jet_purity}
\end{figure}

Figure \ref{fig:jet_purity} shows the fraction of selected GARLIC clusters which are created by photons, or by other classes
of particles: charged and neutral hadrons, electrons and positrons, and others (mostly muons). 
Above around 1 GeV, around 95\% of selected clusters
are produced by photons, with the largest contamination coming from neutral hadrons. As has been explained above,
confusion between photons and neutral hadrons is not particularly important for the measurement of jet energy using
particle flow. The contamination from charged hadrons is at the level of 1-2\% in this energy range. At energies below 1 GeV,
the contamination due to charged hadrons becomes larger, reaching around 18\% at energies below 200 MeV.

\begin{figure}
\center
\includegraphics[width=0.48\textwidth,keepaspectratio]{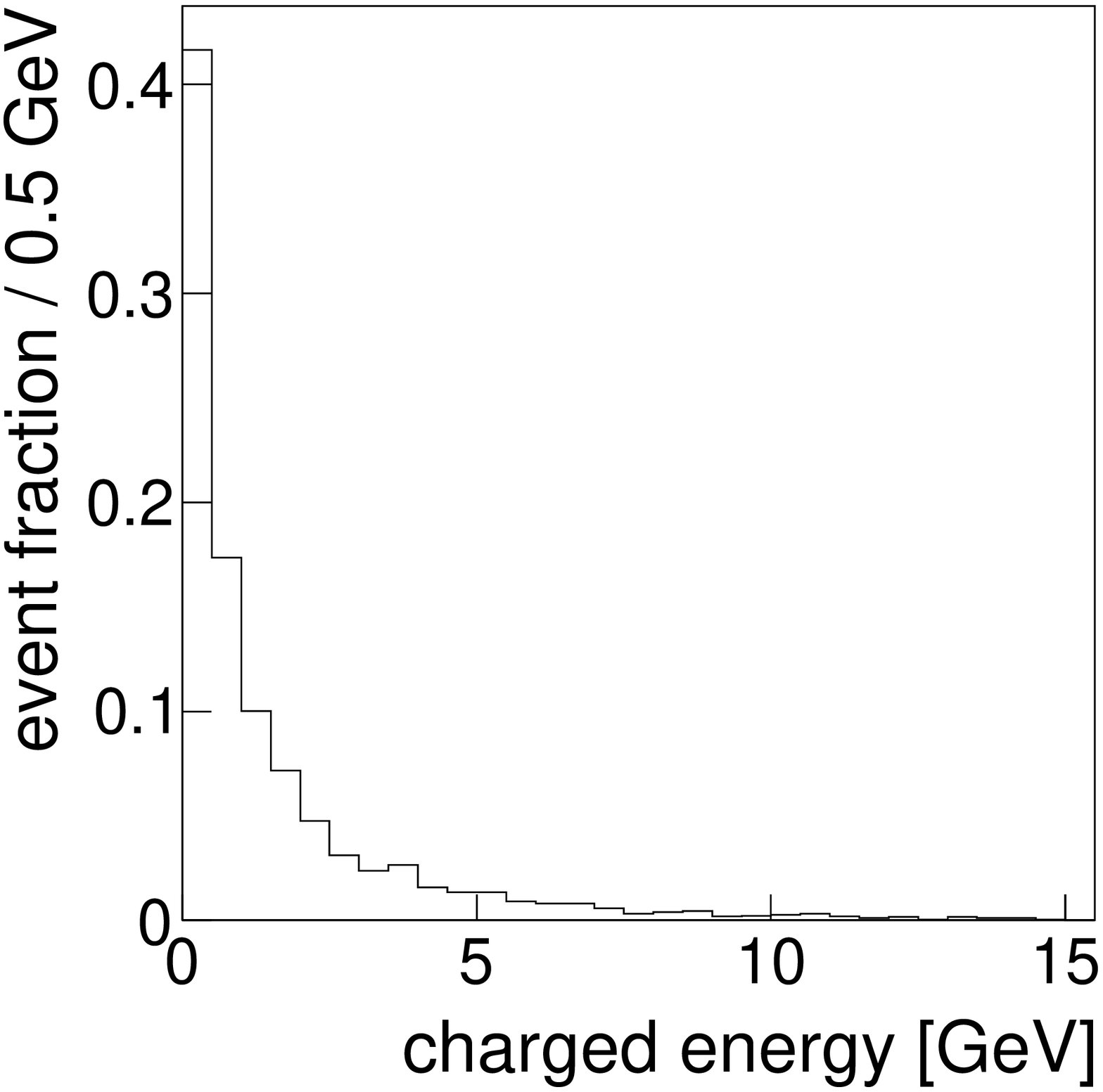}
\includegraphics[width=0.48\textwidth,keepaspectratio]{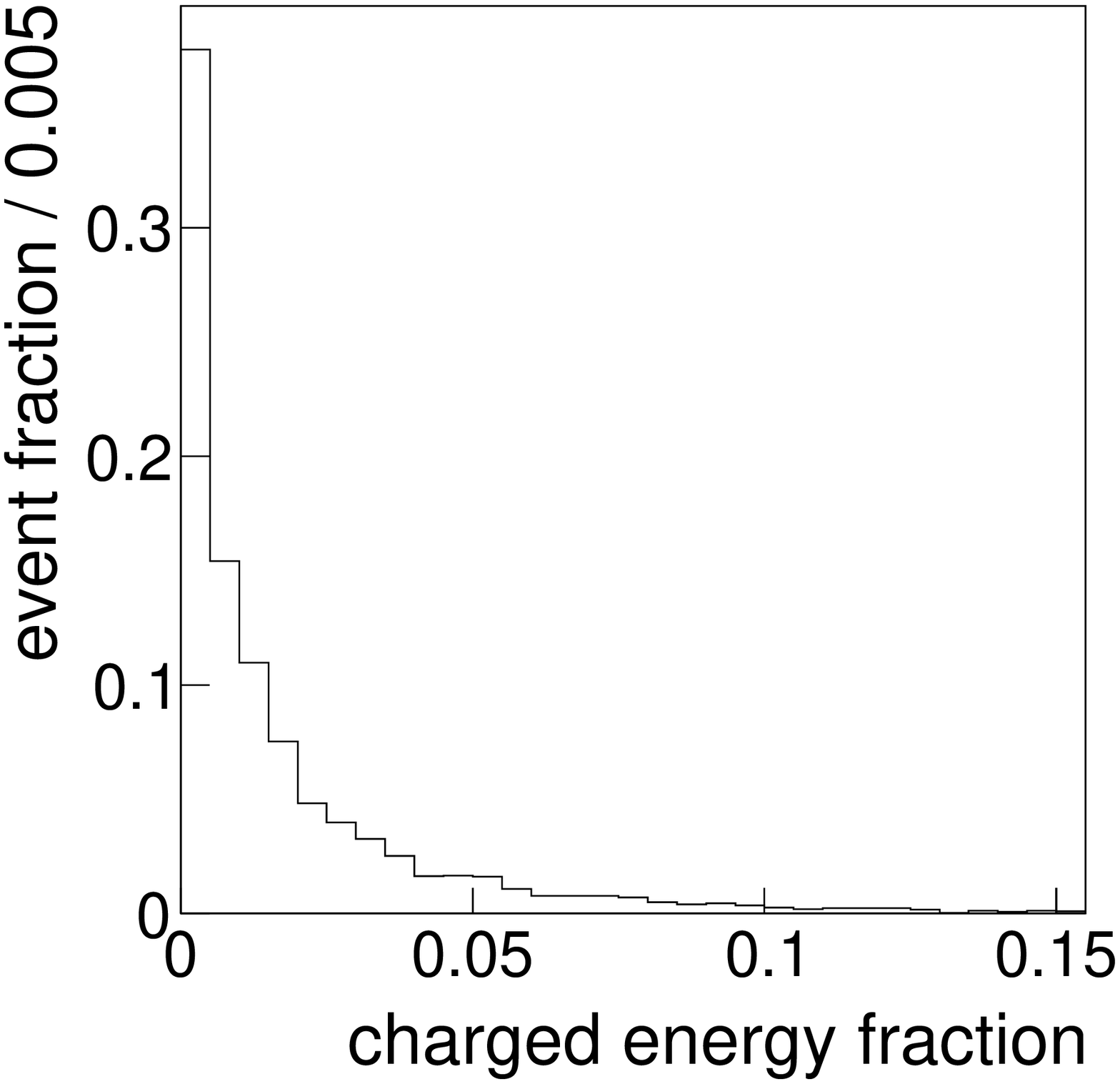}
\caption{Charged hadron contamination in GARLIC clusters, per 4--quark event at 500 GeV.
The left plot shows the absolute charged energy mis-identified, and the right plot the
fraction of the total GARLIC energy it represents.}
\label{fig:jet_toten_contamination}
\end{figure}

Considering the total energy reconstructed by GARLIC in an event, the contamination from charged hadrons amounts to 
around 2\% on average, 0.5\% from electrons and positrons (these two cases of charged particles are the most detrimental 
from the point of view of PFA since they lead to a double counting of energy), 
around 4\% from neutral hadrons (benign from a PFA perspective), and the remainder produced by true photons.
Figure \ref{fig:jet_toten_contamination} shows how much energy per event has been selected by GARLIC but was
deposited by a charged hadron, and the fraction of the total event GARLIC energy which this represents.
This energy is, as expected, strongly peaked towards zero, however it is possible in a small fraction
of events to mis-reconstruct several GeV of energy deposited by charged hadrons, corresponding to several percent of the 
total estimated photon energy. This will contribute to the confusion term of a full PFA.

\begin{figure}
\includegraphics[width=0.45\textwidth,keepaspectratio]{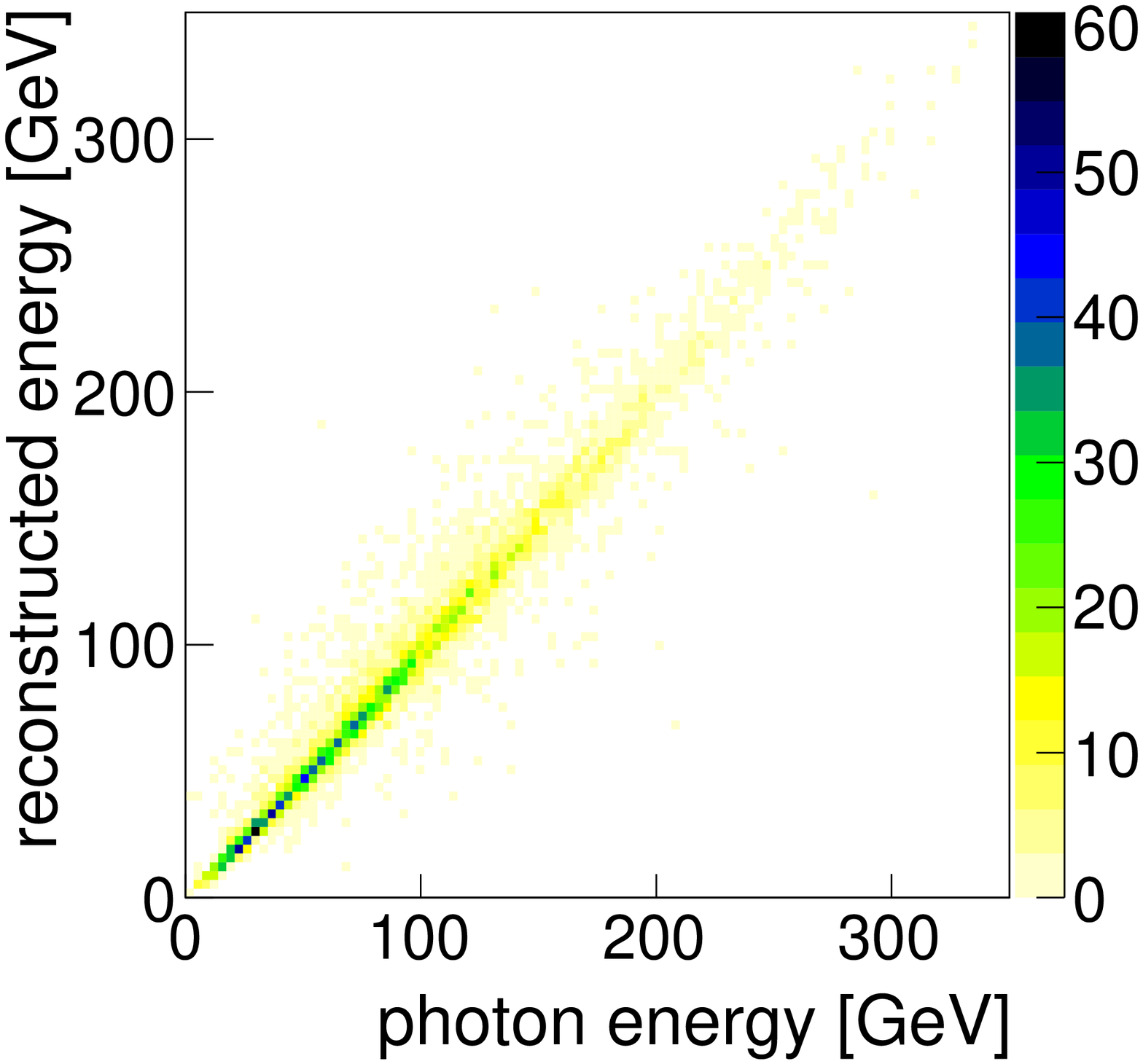}
\includegraphics[width=0.45\textwidth,keepaspectratio]{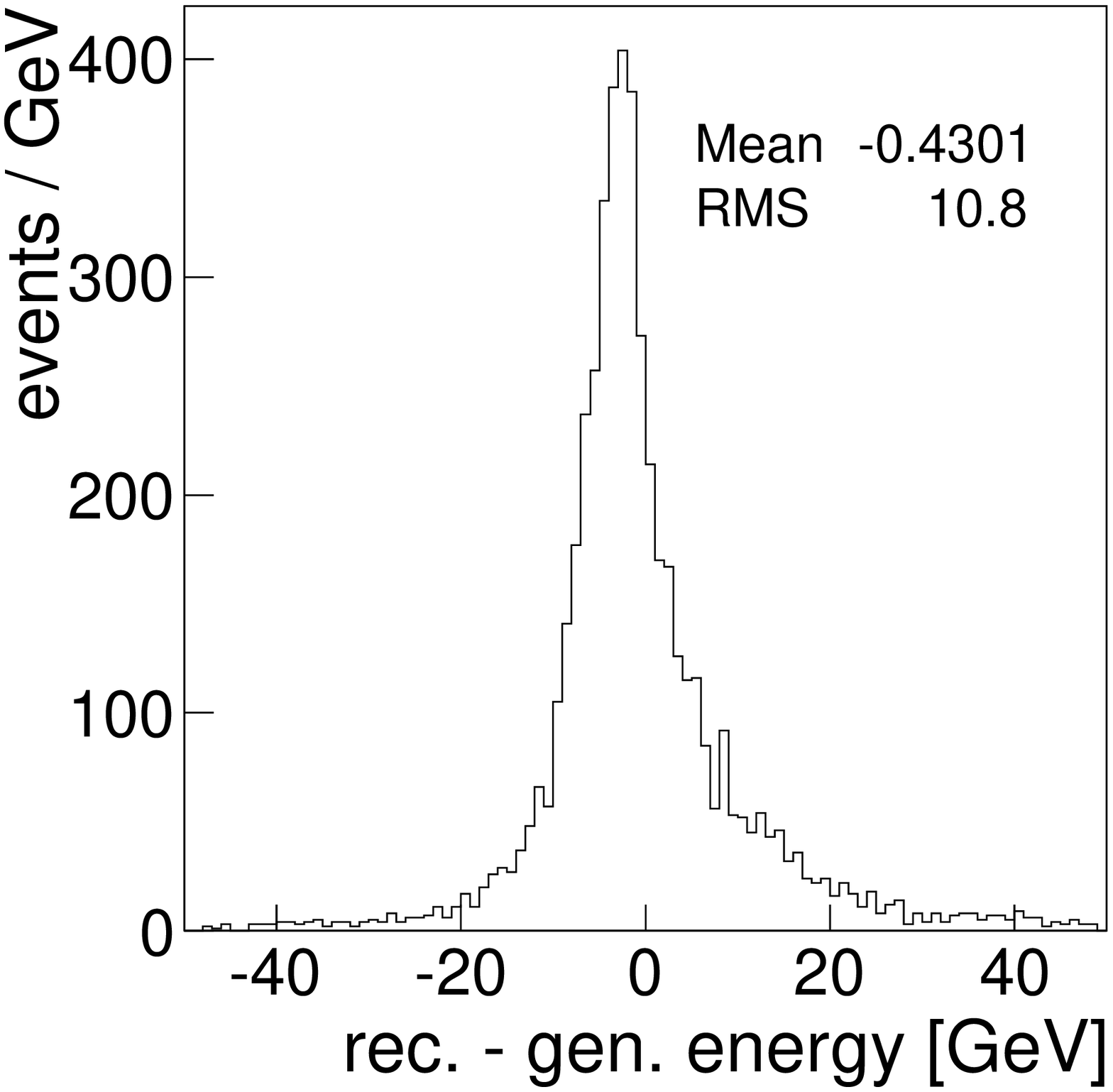}
\caption{Comparison of total generated and reconstructed photon energy per event in 500 GeV four quark events.
The left plot shows the correlation between these two quantities, and the right plot shows the difference between
the reconstructed and generated total photonic energy.}
\label{fig:garlicPerEvent_uddu_500}
\end{figure}

Figure \ref{fig:garlicPerEvent_uddu_500} shows the comparison between the total generated photon energy and
that reconstructed by GARLIC in an event (generated photons are required to lie within the fiducial volume of the detector).
The distribution of the difference in generated and reconstructed energy has two components of similar size,
a narrow central peak with a width of around 4 GeV (representing less than 1\% of the total event energy of 500 GeV), 
and a wider distribution due to cases in which energy deposits were mis-identified.
The resulting distribution has an RMS of around 11 GeV, representing around 2\% of the total event energy, which
is not at a level to dominate the target energy resolution. Since mis-identified energy is largely
due to confusion with neutral hadrons, the effect of this resolution on the total jet energy resolution will be 
smaller than this figure suggests.

Such an detector and reconstruction technique is rather insensitive to dead cells and noise, provided that
they are randomly distributed in the detector. A previous study \cite{pfa_jcb_hv} has demonstrated that
5\% dead sensors, randomly distributed in the ECAL, have a negligible effect on the overall
jet energy resolution. This can understood by noting that each particle shower is sampled
by a large number of detector cells, so the loss of a small number of cells has a 
correspondingly small effect on the measurement of a single particle shower.
Since the occupancy of the ECAL cells in a typical jet event is very small (at the per-mille level), noise in
individual detector cells typically produces low energy, isolated hits which are trivial 
to remove from the event at the earliest stages of data reconstruction.

%

\section{Conclusions}

An efficient separation of photons and charged hadrons in hadronic jets is essential to the physics programme of the 
next linear collider. A fine-grained electromagnetic calorimeter, as envisaged for 
e.g. the International Large Detector concept, is an essential tool to achieve this.

GARLIC is an algorithm which has been developed to reconstruct and identify photons in such a calorimeter.
The efficiency for photon identification depends on the photon energy and the distance to the
closest charged particle at the ECAL. The performance of this algorithm has been measured in 
4-quark events generated at a centre-of-mass energy of 500 GeV.
For photons with an energy of at least 1 GeV and at least 4 cm from the nearest charged particle,
the identification efficiency is above 99\%, while for photons at a distance of 2 cm from a charged
particle at the ECAL front face the efficiency drops to around 90\%. The purity of the selected photon sample
is 95\% above an energy of 2 GeV, and above 90\% at 1 GeV. When considering an entire event,
the energy due to charged hadrons mis-identified as photon-like is around 2\% on average.

The estimate of the total photonic energy per event, even in the challenging case of four jets produced at
500 GeV centre-of-mass energy, 
is accurate to the level of around 2\%, sufficiently good not to dominate the jet energy resolution.

Future studies of GARLIC will include its combination with algorithms designed to reconstruct the hadronic component,
giving a complete particle flow reconstruction.

\acknowledgments

The research leading to these results has received funding from the European Commission under the 
FP7 Research Infrastructures project AIDA, grant agreement no. 262025.

\bibliographystyle{JHEP}
\bibliography{garlicRefs}

\end{document}